**Waldemar Grabski, Wiktor B. Daszczuk**

# PRIORITY RULES ON ATN (PRT) INTERSECTIONS

*In Autonomous Transit Networks some basic elements influence the throughput: network structure, maximum velocity, number of vehicles etc. Other parameters like station structure, dynamic routing or vehicle behavior on intersections play minor role. Yet in highly congested nets, when vehicles interfere in the traffic, some subtle decisions may influence overall system ridership. We tested the impact of intersection priority rules on passenger waiting time, which measures the throughput. The dependence occurred its relevance in a crowded network.*

**INTRODUCTION**

Autonomous Transit Network ATN (Personal Rapid Transit PRT) is a communication means using autonomous guided vehicles traveling on a track separated from urban traffic (typically elevated) [1]. It consists of nodes: stations, capacitors and intersections, interconnected by track segments. We assume two types of segments: road with lower maximum velocity and highway with higher maximum velocity. Moreover, highway segments to dot connect stations nor capacitors, only intersections are allowed. The intersections are of two types: "fork" and "join" ("x" type intersections are not allowed).

A research concentrates on effectiveness of Personal Rapid Transit [2][3][4][5][6]. If a network works in equilibrium conditions [7], the main method of performance improvement is empty vehicle management [5][7][8] (as little may be done with occupied ones: the only option is ride-sharing [9][10][11]). Yet, if the number of vehicles causes congestion, some subtle decision may improve the traffic conditions. We have tested two concepts of increasing the throughput of ATN network: dynamic routing and intersection priority rules.

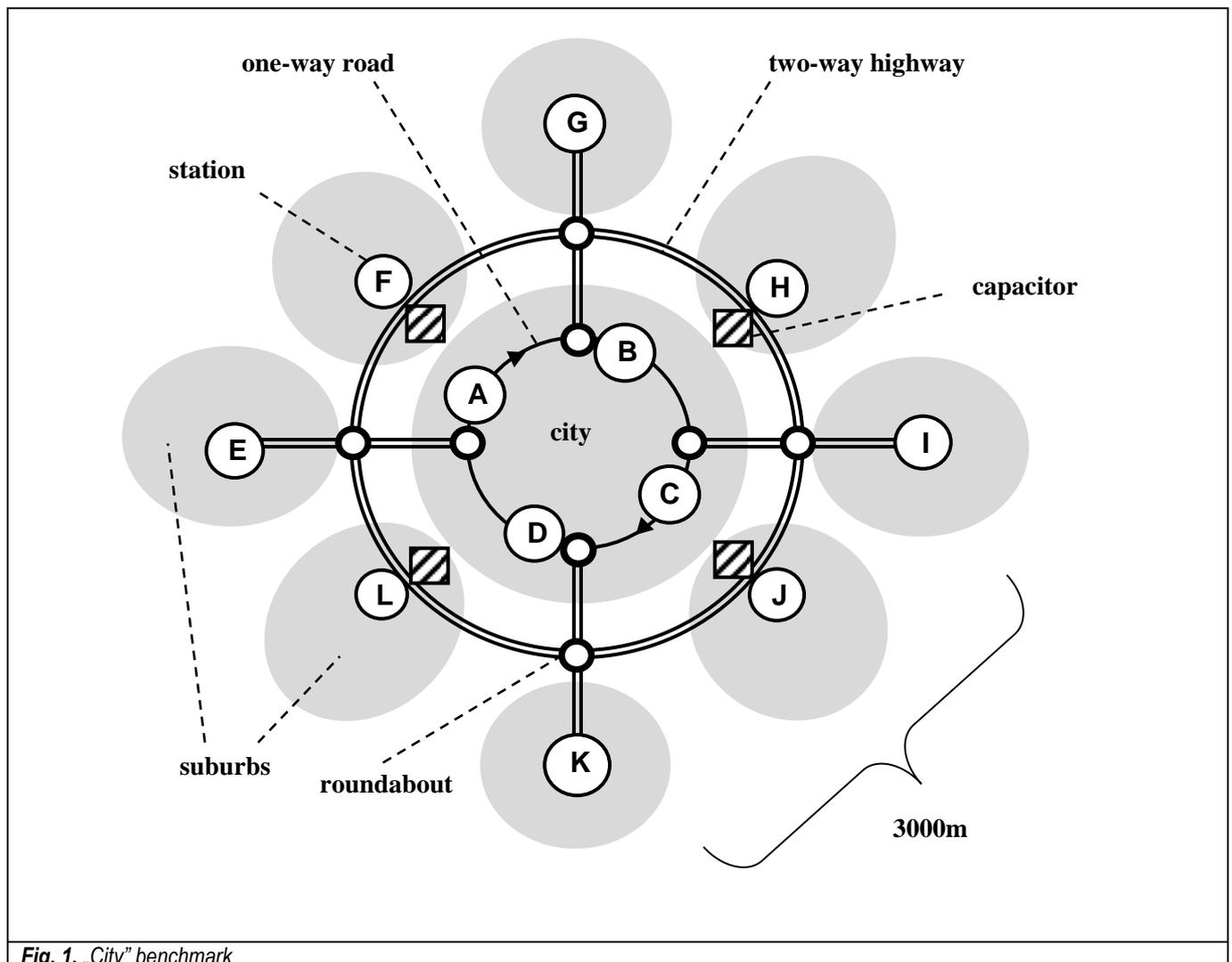

*Fig. 1.* „City" benchmark

The question arises: how to identify congestion conditions? Of course, inserting many vehicles to the network causes they interference and the throughput drops down. But for research we need a single parameter that measures congestion degree, and a single value of this parameter (a threshold) which identifies the network is overcrowded. We decided to use average squared delay for this purpose.

The research on dynamic routing is based on insertion of "Sunday driver" vehicles, which drive slowly and block other vehicles. As the only point of decision in ATN vehicle run is "fork" type intersection, we run the route planning algorithm in front of every such intersection. The Dijkstra planning algorithm [12] is equipped with segment cost function dependent on segment length, segment maximum velocity and segment occupation, everyone with coefficient parameter, The experiments gave inconclusive results, as output parameter (average passenger waiting time) is not monotonic as a function of the segment occupation coefficient.

The second investigation concerns priority rules on "join" intersections. If two vehicles approach an intersection, and if driving with their actual velocities would cause their bringing too close to each other (closer than assumed separation), one of the vehicles must slow down or even stop. Of course, other vehicles that possibly follow this slowing down vehicle must slow down as well. This causes a congestion effect.

We assume the fair rule on intersection of one type segments, if they are both road segments of both highway segments. The fairness is achieved using slider principle (priority assigned alternately to the both segments). The research concerns intersection of different type segments: road and highway. We tested the variants of choosing a vehicle which is not forced to slow down in described situation: on highway, on road or slider principle.

In the research we use the Feniks simulation tool [13], elaborated in Eco-Mobility project [14]. In the project, several simulation-based research programs were carried [7][8][15][16]. The Feniks simulator is still under development, a next step concerns dual-mode operation, in which the vehicle may leave PRT network and continue their trips in ordinary urban traffic.

In Section 1 the benchmark network is presented. Section 2 describes the manner of identification of network saturation point. Simulation results are covered in Section 3. The research is concluded in Section 4.

## 1. THE BENCHMARK

The "City" benchmark [17] is presented in Fig. 1. Double-line segments are two-way highways (two tracks in opposite directions). Single-line segments are ordinary roads. Circles with letters are stations while dashed squares are capacitors. Bold circles are roundabouts allowing to change direction to any of four outcoming track segments.

It is a model of a town, with central city and suburbs. A highway surrounds the city and connects suburbs with each other and with the city. There are 12 stations and 4 capacitors. Total track length is about 33km.

The traffic parameters are:
– 5 berths in every station,
– Highway max velocity 15m/s,
– Road max velocity 10m/s,
– Max acceleration and deceleration 2m/s2,
– Boarding and alighting: triangle distribution with parameters (10, 20, 30) seconds,
– separation 10m,
– Origin-destination matrix a matrix filled with equal values (every trip target chosen randomly).

## 2. IDENTIFICATION OF NETWORK SATURATION POINT

Saturation point is a situation in which the vehicles interfere „disturbing" each other. The disturbing reveals in enlarging of a trip time.

In papers concerning ATN simulation, some measures of traffic quality are used. For example, maximum passenger waiting time or average waiting time is used [18]. Yet, we prefer a single scalar which unambiguously measures the network saturation. We elaborated a synthetic parameter: Average Squared Delay (ASD [7])

$$ASD = \sqrt{\frac{\sum_{full\ trips} \Delta_{ft}^2}{number\ of\ full\ trips}} \quad (1)$$

(square root of sum of squared relative delays of full trips ($\Delta_{ft}$) divided by number of full trips). The delay $\Delta_{ft}$ (in [%]) is counted as actual trip time divided by nominal trip time (with maximal allowed velocity). When the vehicles do not interfere, in typical PRT network ASD is about 15%. A designer may designate a threshold of average squared delay over which the network is assumed to be congested. The results of ASD for a number of vehicles between 48 and 320 is shown in Table 1 and illustrated in Fig. 2. In our example, having the basis of ASD=15%, we may define that the network is congested if ASD rises 25%. This is reached at the number of vehicles about 240.

*Tab. 1.* Identification of network saturation point

| Number of vehicles | Average Squared Delay [%] |
|---|---|
| 48 | 15.29 |
| 72 | 15.71 |
| 96 | 16.35 |
| 120 | 16.87 |
| 144 | 18.08 |
| 192 | 20.28 |
| 240 | 24.77 |
| 280 | 28.76 |
| 320 | 35.97 |
| 192 | 20.28 |

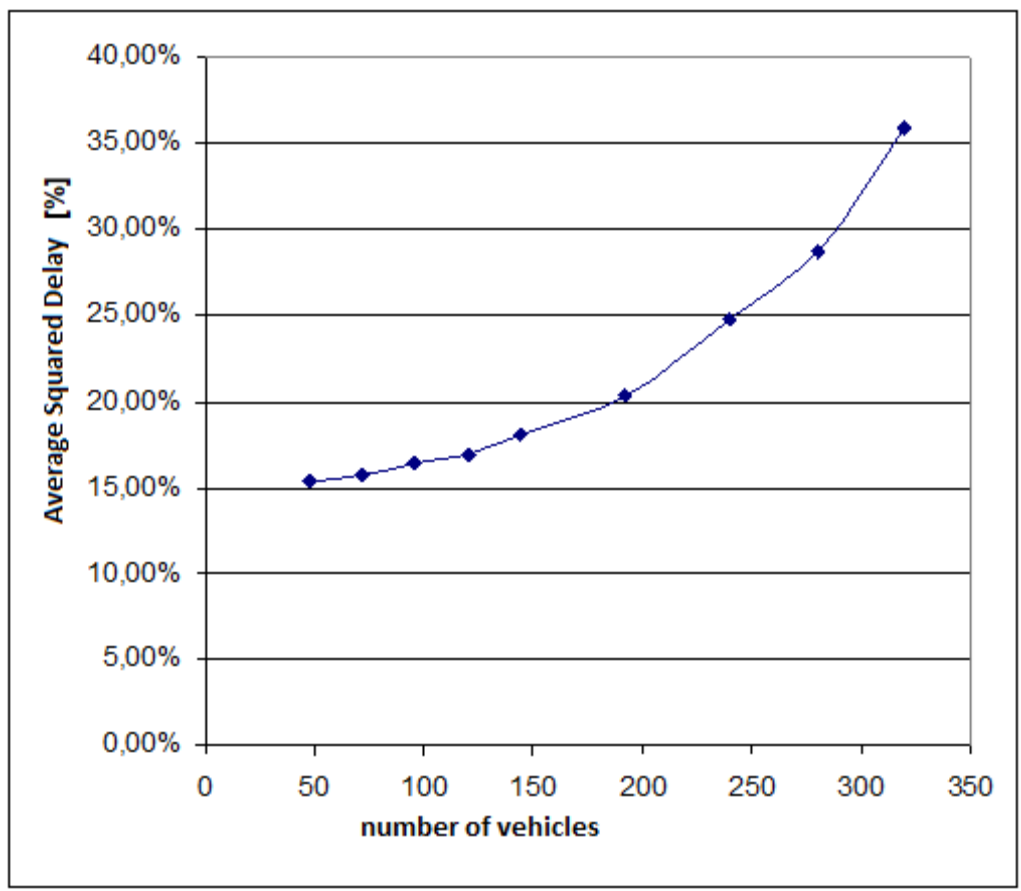

*Fig. 2. Identification of network saturation point*

## 3. SIMULATION RESULTS

For the „City" model (Fig. 1) 240 vehicles were used in simulations (saturation point). The three input intensities were used: 320, 480 and 960 groups/h (one trip is organized for every group of passengers of cardinality 1-4 persons). For every input intensity, all three priority rules were tested:
- Priority for highway,
- Slider principle (alternating priority for highway and for road),
- Priority for road.

The results show influence of priority rules on the throughput of PRT network. The influence rises in the input intensity. It is reasonable because higher input results in more trips to be organized and the actual congestion of the network is higher. Because the differences are not clearly visible for input races 320 and 480 groups/h, they are enlarged in Fig. 4. The plots clearly show, that taking the "priority for road" improves the throughput by 9-30% compared with "priority for highway". Furthermore, "slider" improves by 11-14% compared with "priority for highway". This unintuitive behavior (we expected that the highway should be favored) may be explained such that vehicle queues waiting to cross the intersection are easier to create on the roads because there is a lower maximum speed than on the highway.

## 4. CONCLUSIONS

In the design of a PRT network, many decisions should be made: on a location of stations and capacitors, on the segments category (road/highway), which may differ in curvature and slope, stations size and number of vehicles. In the paper we showed that in the network working in congestion conditions, even subtle traffic rules may tune the throughput of ATN system. We tested priority rules, but station capacity, dynamic routing (with other algorithm than ours) and similar mechanisms may also improve the quality.

**Zasady pierwszeństwa na skrzyżowaniach sieci ATN(PRT)**

*W sieciach transportowych ATN (Autonomous Transit Network) na przepustowość mają wpływ jej główne elementy jak struktura sieci, maksymalna prędkość, liczba pojazdów itd. Inne parametry jak struktura przystanków, dynamiczny*

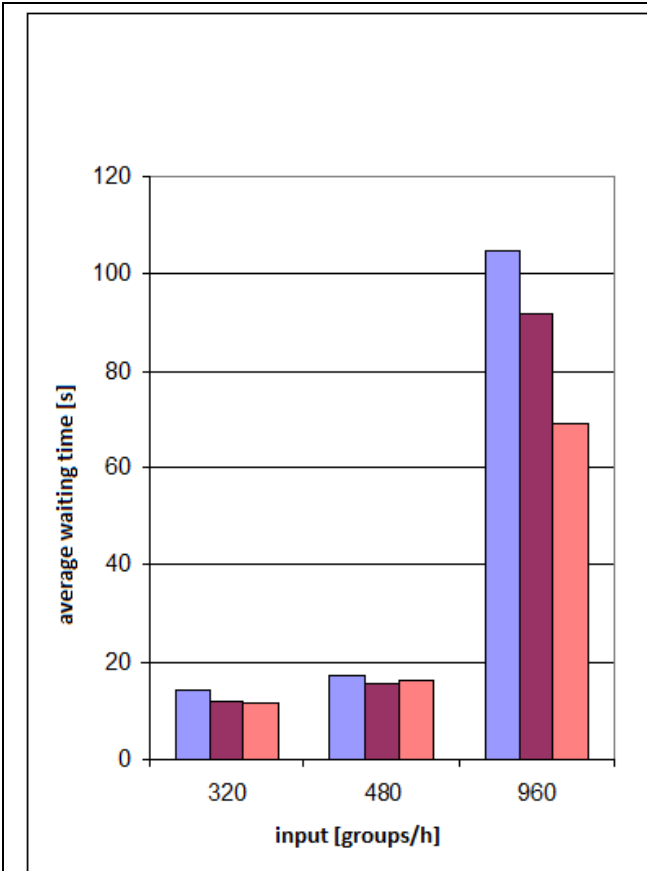

**Fig. 3.** *The effect of priority rules on the „join" intersection to average waiting time, with different input rates*

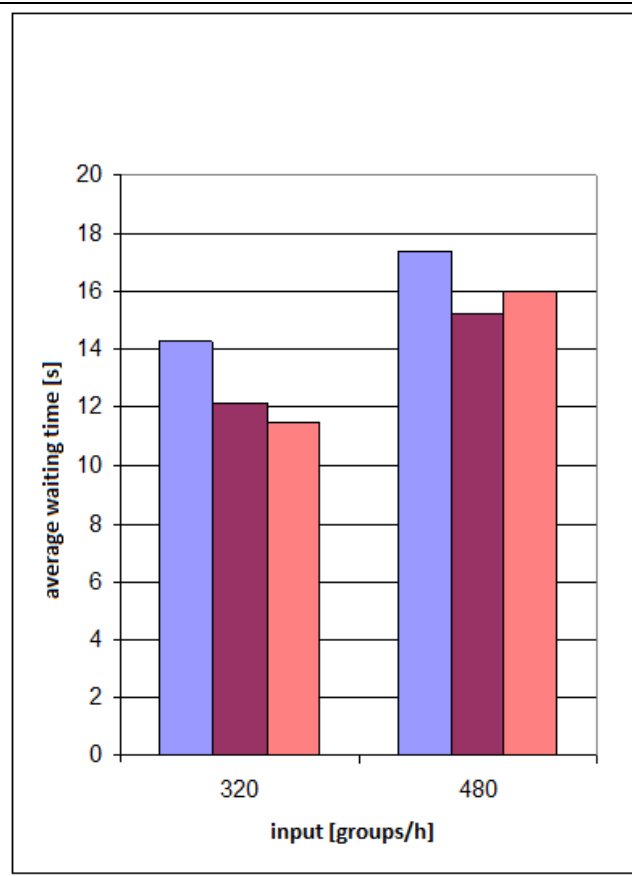

**Fig. 4.** *Plot from Fig. 3, except for the case of the largest input*

*4*

*wybór drogi czy zachowania pojazdów na skrzyżowaniach grają niewielką rolę. Jednak przy dużym zagęszczeniu pojazdy przeszkadzają sobie wzajemnie i pewne subtelne decyzje mogą wpłynąć na zdolność przewozową sieci. Zbadaliśmy wpływ zasad pierwszeństwa na skrzyżowaniach na czas oczekiwania na pojazdy, który jest miarą przepustowości. Zależność okazała się istotna w zatłoczonej sieci.*

Authors:

**Waldemar Grabski**, MSc – Warsaw University of Technology, Institute of Computer Science, Nowowiejska str. 15/19, 00-665 Warsaw, wgr@ii.pw.edu.pl

**Wiktor B. Daszczuk**, PhD – Warsaw University of Technology, Institute of Computer Science, Nowowiejska str. 15/19, 00-665 Warsaw, wbd@ii.pw.edu.pl